\documentclass[12pt,preprint]{aastex}

\slugcomment{Submitted to ApJ Letters, 17/08/01}

\shorttitle{LG satellites}
\shortauthors{Prada and Burkert}

\begin{document}


\title{The Fundamental Line of the Local Group Satellites}

\author{Francisco Prada}
\affil{Centro Astron\'omico Hispano-Alem\'an, Apdo 511, E-04080 
Almer\'{\i}a, Spain}
\authoremail{prada@caha.es}
\and
\author{Andreas Burkert}
\affil{Max-Planck-Institut f\"{u}r Astronomie, K\"{o}nigstuhl 17, D-69117 
Heidelberg, Germany}
\authoremail{burkert@mpia-hd.mpg.de}

\begin{abstract}

We present a new correlation between the mass-to-light ($M/L$) ratio and
the mean metallicity for the satellites of the Local Group. This
relation together with their central surface brightness 
define a Fundamental Line where metal poor and low surface brightness
dwarfs are dark matter dominated while metal rich high surface
brightness systems will have a low $M/L$ ratio. This Fundamental
Line is independent of distance to the central galaxy (M31 and the
Milky Way) and morphological type among other global parameters as
their star formation history. The new $M/L$-metallicity 
relation indicates that dwarf spheroidal galaxies are dark matter confined. 
It can be interpreted if satellites experienced a continuous
loss of metals with a final episode of gas ejection at the end of
the star formation epoch. Only Globular Clusters are found to lie outside
the Fundamental Line of dwarf satellites. Unlike satellite galaxies they
are not dark matter confined.

\end{abstract}


\section{Introduction}

The local group (LG) consists of two dominant giant spirals and a system of 
satellites.  Due to the small distances this sample of well studied dwarf 
galaxies provides unique insight into the structure and evolution of 
low-luminosity systems and their tidal interaction with the parent galaxy.
Furthermore their distribution and dynamics gives an excellent opportunity 
to probe the general paradigm of galaxy formation on galactic 
scales (see Klypin ${\it et \, al.}$ 1999, and references therein).

The LG satellites are part of a large zoo of dwarf galaxies which
by number are the dominant population in the universe (e.g. Ferguson and
Binggeli 1994).  Like in other dwarf galaxies, well-defined
correlations have been found between their measured properties, like
absolute magnitude, surface brightness, characteristic radius and
metallicity (Armandroff ${\it et \, al.}$ 1998; Caldwell 1999). Their 
low surface brightnesses and metallicities which correlate well with 
luminosity have been interpreted as a result of substantial gas loss in 
the early stages of evolution (Larson 1974; Vader 1986; 
Arimoto and Yoshii 1987, Dekel and Silk 1986).  In this 
case one might expect that star formation stopped in dwarf galaxies after 
a short initial period due to the removal of the gas in a wind. This is however
not confirmed by recent observations. As summarized e.g. by Da Costa
(1998) and Grebel (2000) (see also Da Costa ${\it et \, al.}$ 2000 for recent
observations of dwarf spheroidal galaxies around M31) the stellar
populations of the LG satellites show a fascinating diversity
of star formation histories. Several systems are old.  There exist
however also systems with intermediate or young stellar populations
and those which experienced several distinct epochs of efficient star
formation with long periods of quiescence. As pointed out e.g. by van
den Bergh (1994) the fraction of 
intermediate-age population stars correlates with galactocentric distance, 
indicating that satellite galaxies do not evolve in isolation but are 
affected by interaction with their parent galaxy.

Especially the smallest galactic systems, the dwarf spheroidals
(dSph) have very large mass-to-light ratios ($M/L$) which appear to
anticorrelate with their intrinsic brightness, indicating the presence
of a dark matter component with a constant mass and a variable luminosity
(Mateo 1998). This result is very interesting as it might provide
important information on the dark substructure of galaxies and maybe even
on the nature of dark matter (e.g. Burkert 2000).  An alternative scenario
is however that the dSphs with large apparent velocity dispersions are
simply coherent but unbound groups of stars on similar orbits (Kuhn 1993,
Kroupa 1997, Klessen and Kroupa 1998).  In this case, these systems might
not contain large amounts of dark matter (Moore 1996, Burkert 1997).  This 
result would be interesting as it indicates that the satellites do not 
represent the typical subunits out of which their parent galaxies 
formed but rather formed during the merging of an LMC-type galaxy which broke
into pieces as a result of its tidal interaction with the Local Group.

This Letter combines recent detailed observations of the Local Group
satellite galaxies and presents a puzzling new correlation between
their metallicity and their mass-to-light ratio. We argue that this
correlation could be understood if these systems experienced efficient
loss of metals during the star formation phase with a final episode
of gas loss at the end of the star formation epoch. It also indicates
that dwarf spheroidals are gravitationally
bound and contain a large amount of dark matter.

\section{The M/L - metallicity - ${\Sigma}$ relation of LG satellites}

 In total there are 28 dwarf galaxies within a volume of
radius 570 kpc ($400h^{-1}$ kpc assuming a Hubble constant $h=0.7$,
 $H_o = 100h$ km s$^{-1}$ Mpc$^{-1}$) centered on
the Milky Way and M31. They have V-band absolute magnitudes between
$-19 < M_V < -8.5$ and can be divided into few different morphological types: 
dSph, dwarf ellipticals (dE), gas rich spirals and irregulars (S,Irr,dIrr) 
and transition dwarfs (dSph/dIrr) (see Grebel 2000,2001). The median 
distance of the LG satellites to the central galaxy is 
$90h{^{-1}}$ kpc with the dSph being concentrated around the two 
giant spirals. 

Values and errors for the mean metallicity ($[$Fe/H$]$) of the dominant 
intermediate/old population, the V-band central surface brightness 
(${\Sigma}_{0,V}$) and the V-band total Luminosity ($L_V$) were taken from 
the most recent compilations by Grebel (2000) and van den Bergh (2000). Data 
for the mass are collected from Mateo (1998) for most of the 
satellites and from van den Bergh (2000) for the LMC, SMC, Sculptor and
LGS3. The $M/L_V$ for AndII was found in C\^{o}t\'{e} et al. (2001).

In Figure 1, we show the relation between the V-band mass-to-light ratio 
($M/L_V$, here $M$ refers to the dynamical mass, see Mateo 1998) and
the mean metallicity for our sample of Local Group satellites. 
The errors for $[$Fe/H$]$ indicate the intrinsic spread of metallicity for
each galaxy (see Grebel 2000). An error of $40\%$ in the
determination of $M/L_{V}$ for our satellites (see Mateo 1998) was
adopted. The $M/L_V$ - metallicity relation is well described by a linear 
law of slope unity (solid line in Figure 1)

\begin{equation}
log \frac{M}{L_V} = -0.4 -[Fe/H]
\end{equation}

with a Chi-square error
($\chi^2$) of 1.4, i.e. satellites with high $M/L_V$ have low metallicities. 
This relation is independent of distance to the central galaxy and 
morphological type among other global parameters as their star formation 
history. 

We also display in Figure 1 the $M/L_V$ ratio versus the V-band central 
surface brightness for the same galaxies (except M32 
due to its very bright nucleus and Phoenix and Pegasus for which no 
${\Sigma}_{0,V}$ values were found in the literature). The solid line is 
a linear fit to the data with slope 
$\Delta log(M/L_V)/\Delta{\Sigma}_{0,V} = +0.27$ and $\chi^2=1.30$
(see also Armandroff et al. 1998).

The scatter in the M/L - metallicity relation improves when we include
as a second parameter the central surface brightness.  A Principal Component 
Analysis shows that $93\%$ of the variance of the data are in a line. A 
least-squares fit yields the following log-linear regression with 
a $\chi^2=1.05$:

\begin{equation}
log (M/L_{V}) = - 0.56 \, [Fe/H] + 0.13 \, \Sigma_{0,V} -2.86
\end{equation}

This relation can be understood as a Fundamental Line for the satellites
of the Local Group (see Figure 2) where the values for the central surface 
brightness and metallicity of a given system determine its $M/L$ ratio. Hence
 metal poor and low surface brightness (LSB) dwarfs are dark matter
dominated systems while high surface brightness (HSB) metal rich dwarfs 
will have a low $M/L$ ratio. In addition, for a given surface brightness, 
systems with low metallicities have large mass-to-light ratios. Besides M32 
which is peculiar in many respects we did not find any system located 
outside this relation, even if we include the data for those few dwarfs 
outside our volume and located in the outer fringes of the Local Group 
(see van den Bergh 2000).

\section{Discussion}

In the previous section we have shown that all local group satellites follow
a fundamental line that combines the mass-to-light ratios with their 
metallicity and central surface brightness. We therefore conclude that dSphs,
like the other LG galaxies are  unlikely to be
chance projections of disrupted galaxies but instead are strongly dark 
matter confined. If they actually were tidal streams seen in projection it 
would be difficult to understand why the projected velocity dispersion which 
depends on the orbit and the projection angle should follow the sample tight 
relationship as the other galaxies of the LG.

In fact we can now formulate a more general question: do there exist yet 
undetected high-surface brightness metal poor dwarfs with high M/L ratios 
or low-surface brightness metal rich dwarfs with low M/L ratios? These 
systems would not follow the Fundamental Line relation and might have formed 
by a different physical process than tyical satellites. The 
Globular Clusters (GC) around M31 and the Milky Way are systems without
dark matter ($M/L=1-3$), unlike galaxies, with a large range in metallicities
from metal poor to metal rich clusters. They therefore lie outside of the 
Fundamental Line of satellites. There are however also a few cases of GC that 
follow our $M/L_V$ - [Fe/H] relation. Mayall II (G1) in M31 and Pal 13 in 
the Milky Way have anomalous $M/L_{V}$ ratios (7.5 and 20 respectively)
with a spread in metallicity ($[Fe/H]$=-0.98$\pm$0.45 and
-1.67$\pm$0.20 respectively) compared to the majority of globular clusters 
which are single-population systems without such a spread
(see Meylan 2001 and references therein). This supports the suggestion that 
at least these 2 systems with high $M/L$ ratios might actually be the 
remaining cores of dwarf elliptical galaxies instead of real GCs.

The chemical enrichment model of Dekel and Silk (1986) might explain the 
observed correlation between ($M/L_V$) and $[Fe/H]$. To demonstrate this 
we adopt a universal initial baryon-to-dark matter mass fraction 
$\eta_b=M_g/M_{DM} \approx 0.1$, where $M_g$ and $M_{DM}$ is the gas and 
dark matter mass, respectively. We assume that the satellites formed through 
a first period of star formation which might have had several epochs of 
quiescence and which ended when the rest of the gas was ejected either by 
a strong galactic wind and/or through tidal stripping. During the star 
formation epoch it is likely that hot metal-enriched gas was also lost 
in a continuous wind. As demonstrated by Mac Low and Ferrara (1999) 
supernova driven galactic winds could  blow out a large fraction of the 
newly produced metals without removing much gas or destroying the galaxy. 
If we define the effective yield $y_{eff}$ as the mass fraction of metals 
that were produced by newly formed stars and that were not ejected in the 
wind, the mean iron abundance of the stellar system can be approximated 
for low metallicities as

\begin{equation}
Z = y_{eff} \frac{M_*}{M_g} 
\end{equation}

\noindent Here, Z is the mean  iron mass fraction and $M_*$ is the stellar 
mass. Adopting a solar metallicity of $Z_{\odot}=0.02$ and with 
$[Fe/H]=log(Z/Z_{\odot})$ we find

\begin{equation}
log (Z/Z_{\odot}) = log \left( \frac{y_{eff} (M_*/L_V) }{0.02 \eta_b} \right)  - [Fe/H]
\end{equation}

\noindent This relation agrees with equation (1) if 

\begin{equation}
y_{eff} = 0.008 \eta_b \frac{M}{L_V} \approx 0.001
\end{equation}

For a typical initial mass function the expected iron yield is 0.04 as 
inferred e.g. from the most metal-rich stellar systems which on average have 
metallicities of twice the solar metallicity. This yield could be reduced 
roughly by a factor of 3 if supernovae type Ia did not contribute.
However, even in this case, the yield would be an order of magnitude too 
large. This problem can be solved if we postulate a continuous outflow of 
hot metal-rich gas during the star formation phase in satellite galaxies 
which removed 90\% of the iron with only 10\% being retained. 

In summary, the observed correlation between the mass-to-light ratio and 
the metallicity of dwarf satellites can be explained by a simple chemical 
model of continuous loss of metals with a final gas ejection phase at the 
end of the star formation epoch. Although this approach is promising it is 
not clear why the fraction of metals ejected in the wind should be 
independent of the total mass or surface density. More sophisticated models 
will be required to understand the origin of the fundamental line in details.

\acknowledgments

We are grateful to E.Ofek, C.Pryor, O.Valenzuela, 
A.Ferrara, B.Moore, R.Ibata, E.Grebel, J.Gallagher and 
other participants of the Ringberg workshop on Low-mass dwarf galaxies 
for discussions and comments. F.P. thanks the hospitality of the 
MPIA were part of this work was done and always to Anatoly Klypin for many
fruitful discussions.

\clearpage


\begin{figure}
\plotone{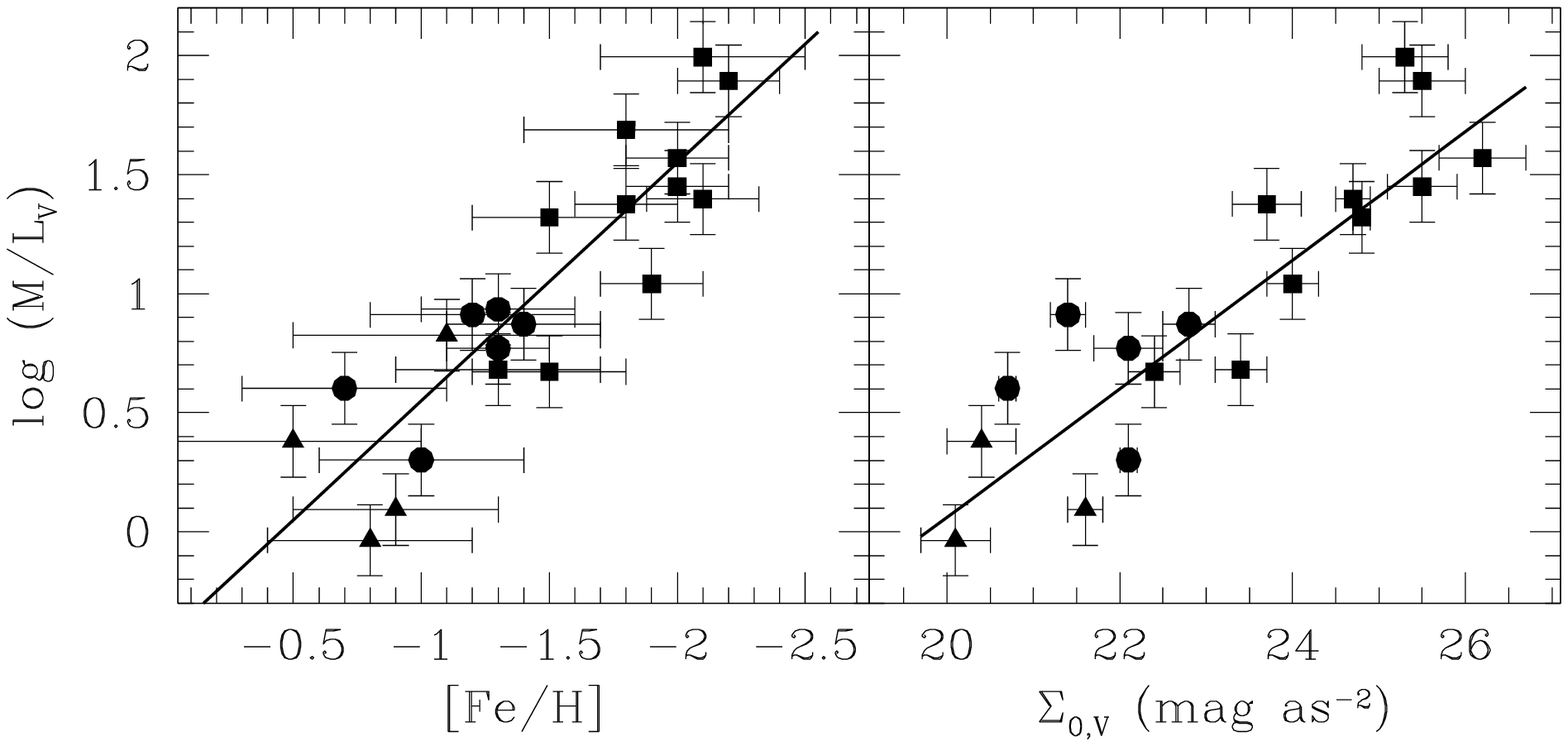}
\caption{{\it Left:} Relation between the  mass-to-light ratio ($M/L_V$) and
the mean metallicity for our sample of Local Group 
satellites. Filled square symbols are the dSph and
transition objects (dSph/dIrr); filled circles are
S, Irr and dIrr and the filled triangles are
the dE satellites. The solid line represents a single
linear law.{\it Right}: The $M/L_V$ ratio versus the
V-band central surface brightness for the same galaxies. Solid
line is a fit to the data (see text).}
\end{figure}

\clearpage 

\begin{figure}
\plotone{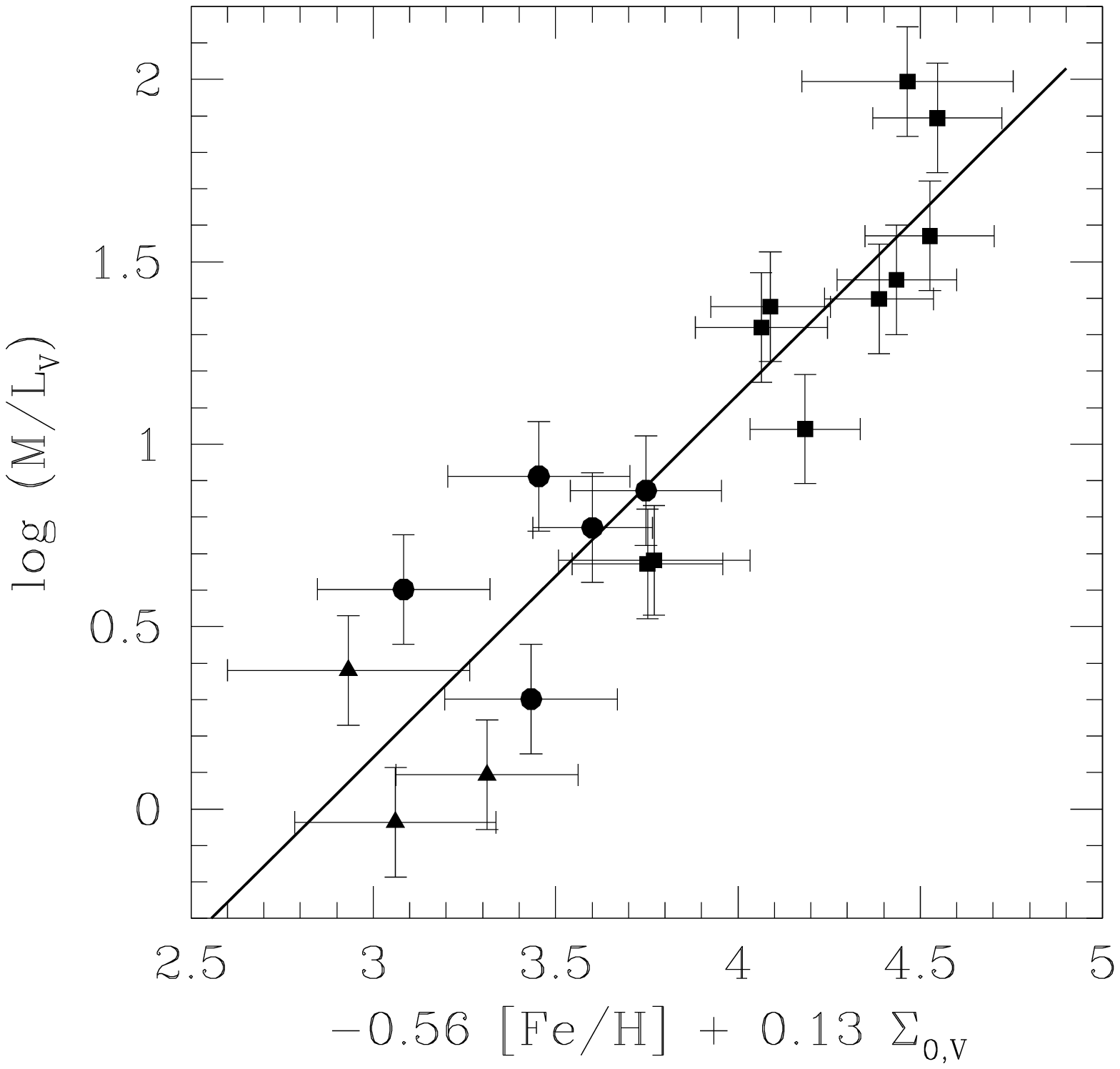}
\caption{Plot of the Fundamental Line of the Local Group dwarf
satellites. Symbols as in Figure 1 (see text).}
\end{figure}


\begin{thebibliography}{}
\bibitem[]{b} Arimoto, N.~\& Yoshii, Y.\ 1987, \aap, 173, 23 
\bibitem[]{b} Armandroff, T.~E., Davies, J.~E., \& Jacoby, G.~H.\ 1998, \aj, 116, 2287
\bibitem[]{b} Burkert, A.\ 1997, \apj, 474, L99
\bibitem[]{b} Burkert, A.\ 2000, \apj, 534, L143
\bibitem[]{b} Caldwell, N.\ 1999, \aj, 118, 1230
\bibitem[]{b} C{\^ o}t{\' e}, P., Mateo, M., Olszewski, 
E. W., \& Cook, K. H.\ 1999, \apj, 526, 147
\bibitem[]{b} Da Costa, G.~S.\ 1998, 
Stellar astrophysics for the local group: VIII Canary Islands Winter School 
of Astrophysics, Edited by A. Aparicio, A. Herrero, and F. Sanchez. 
Cambridge ; New York : Cambridge University Press, 1998., p.351 351
\bibitem[]{b} Da Costa, G.~S., Armandroff, T.~E., 
Caldwell, N., \& Seitzer, P.\ 2000, \aj, 119, 705
\bibitem[]{b} Dekel, A., Silk, J.\ 1986, \apj, 303, 39 
\bibitem[]{b} Ferguson, H.~C.~\& Binggeli, B.\ 1994, \aapr, 6, 67 
\bibitem[]{b} Grebel 2000, in "Star Formation
    from the Small to the Large Scale", Proc. 33rd ESLAB Symp, Noordwijk 1999
    (ESA), eds. F.Favata, A.A. Kaas, \& Wilson
\bibitem[]{b} Grebel 2001, in "Dwarf galaxies and their environment", Bad
    Honnef 2001 (Germany), eds. K.S.de Boer, R.-J.Dettmar, \& U.Klein
\bibitem[]{b} Klessen, R.~S.~\& Kroupa, P.\ 1998, \apj, 498, 143
\bibitem[]{b} Klypin, A., Kravtsov, A.~V., Valenzuela, O., \& Prada, F.\ 1999, \apj, 522, 82
\bibitem[]{b} Kroupa, P.\ 1997, New Astronomy, 2, 139 
\bibitem[]{b} Kuhn, J.~R.\ 1993, \apjl, 409, L13 
\bibitem[]{b} Larson, R.~B.\ 1974, \mnras, 169, 229
\bibitem[]{b} Mac Low, M.-M., Ferrara, A. \ 1999, \apj, 513, 142
\bibitem[]{b} Mateo, M. L.\ 1998, \araa, 36, 435
\bibitem[]{b} Meylan, G. 2001, in "Extragalactic Star Clusters",
 IAU Symposium Series, Vol. 207, 2001, edited by E. Grebel, D. Geisler, and 
D. Minniti
\bibitem[]{b} Moore, B.\ 1996, \apj, 461, L13
\bibitem[]{b} Vader, J.~P.\ 1986, \apj, 305, 669
\bibitem[]{b} van den Bergh, S.\ 1994, \apj, 428, 617 
\bibitem[]{b} van den Bergh, S.\ 2000, 
The galaxies of the Local Group. Published by 
Cambridge, UK: Cambridge University Press, 2000 Cambridge Astrophysics 
Series Series, vol no: 35
\end{thebibliography}
\end{document}